\documentclass[superscriptaddress,floatfix,twocolumn,amsmath,amssymb]{revtex4-2}

\usepackage{lipsum}
\usepackage{graphicx}
\usepackage{xcolor}
\usepackage{mathrsfs}
\usepackage{mathtools}
\usepackage{braket}
\usepackage[normalem]{ulem}
\usepackage{url}
\usepackage{hyperref}
\usepackage[capitalise]{cleveref}
\UseRawInputEncoding

\newcommand{\Tr}{\mathrm{Tr}}

\begin{document}

\begin{abstract}
Controlling quantum systems under correlated non-Markovian noise, particularly when strongly coupled, poses significant challenges in the development of quantum technologies. Traditional quantum control strategies, heavily reliant on precise models, often fail under these conditions. Here, we address the problem by utilizing a data-driven graybox model, which integrates machine learning structures with physics-based elements. We demonstrate single-qubit control, implementing a universal gate set as well as a random gate set, achieving high fidelity under unknown, strongly-coupled non-Markovian non-Gaussian noise, significantly outperforming traditional methods. Our method is applicable to all open finite-dimensional quantum systems, regardless of the type of noise or the strength of the coupling.
\end{abstract}

\title{Quantum control in the presence of strongly coupled non-Markovian noise}

\author{Arinta Auza}
\address{Quantum Photonics Laboratory and Centre for Quantum Computation and Communication Technology, RMIT University, Melbourne, VIC 3000, Australia}

\author{Akram Youssry}
\address{Quantum Photonics Laboratory and Centre for Quantum Computation and Communication Technology, RMIT University, Melbourne, VIC 3000, Australia}

\author{Gerardo Paz-Silva}
\address{Centre for Quantum Dynamics and  Centre for Quantum Computation and Communication Technology, Griffith University, Brisbane, Queensland 4111, Australia}

\author{Alberto Peruzzo}
\address{Quantum Photonics Laboratory and Centre for Quantum Computation and Communication Technology, RMIT University, Melbourne, VIC 3000, Australia}
\address{Qubit Pharmaceuticals, Advanced Research Department, Paris, France}

\maketitle

\section{Introduction}
\label{introduction}
Quantum technology \cite{leuchs2019quantum} represents a cutting-edge frontier in scientific and technological innovation, leveraging the principles of quantum mechanics to revolutionize computation, communication, and sensing. 
Noise in quantum systems, particularly of the non-Markovian and non-Gaussian types, poses significant challenges to the advancement of quantum technologies. When quantum systems interact with complex environments that have memory effects, they experience non-Markovian noise which is troublesome because of the difficulty of expressing the dynamics mathematically, rendering the process of system identification and control challenging. In current quantum computing platforms, while individual quantum gate operations may achieve high fidelity, the cumulative effect of non-Markovian noise across long gate sequences can severely impact overall system performance \cite{White_2020}. This highlights the critical need for addressing strong system-environment interactions to enhance the reliability and efficiency of quantum technologies.

Quantum control techniques \cite{Dong_2010, Dong_2021} are essential for realizing the full potential of quantum technologies, targeting different objectives including state manipulation, noise reduction, and operation optimization.
Quantum control involves optimizing a cost function $C(u, \Theta; G)$, such as gate fidelity, relying on measurable outputs, control inputs $u$, and model parameters $\Theta$ to achieve a target $G$. Since $C$ is an unknown non-linear function of $\Theta$, assumptions are necessary to build a model and derive an estimated cost function $\hat{C}(u, \Theta; G)$. This model-based, or ``whitebox'', approach requires well-defined model parameters, such as Hamiltonian coefficients and noise correlation functions, which are often unknown a priori. Hence, phenomenological observations or assumptions about system states, like a high temperature thermal state of the bath, are used to construct the model.

However, this approach has limitations in achieving high fidelity gates due to its dependency on these assumptions. To address this, data-driven methods like Hamiltonian learning (HL) \cite{Wang_2018, wang2017experimental} and quantum noise spectroscopy (QNS) \cite{norris2016qubit, pazsilva2017multiqubit, krzywda2019dynamical, sung2021multi}  have been developed. These techniques adjust the parameters according to the system response to a selected set of controls to determine the Hamiltonian coefficients and noise correlations, improving the consistency of $\hat{C}(u, \Theta, G)$ with experimental data and enhancing gate fidelity. 
Nevertheless, the effectiveness of these whitebox strategies depends on basic assumptions, e.g., fixed order perturbation theory must be consistent with observed dynamics. Therefore, they are not effective in strong coupling regimes where perturbation theory is inadequate since the complexity of perturbation theory grows quickly with perturbation order, highlighting their inherent limitations.

To address the challenges of whitebox models machine learning techniques have been proposed, including supervised learning (blackbox models) \cite{flurin2020using, papivc2022neural, wise2021using, palmieri2021multiclass, ostaszewski2019approximation, khait2022optimal, zeng2020quantum}, and model-free methods like reinforcement learning (RL) \cite{sivak2022model,giannelli2022tutorial} and direct numerical optimization \cite{khaneja2005optimal, PhysRevA.99.052327, Yang_2020}. These approaches prove particularly useful in complex scenarios, such as many-body systems \cite{leung2022observation}, where creating an accurate model is challenging due to inherent system complexities. However, these approaches do not provide any insight about the system, and model-free methods requires real-time system access, complicating their implementation.

A promising alternative is the ``graybox'' (GB) approach, blending elements of both whitebox and blackbox models. This hybrid method ~\cite{youssry2021noise, youssry2020characterization, youssry2020modeling, youssry2023experimental, perrier2020quantum}, utilizes well-understood model components alongside neural networks to handle uncertainties. This strategy allows for a flexible adjustment between model-driven and data-driven elements based on the available knowledge and training limitations. Unlike purely target-driven models, the GB method provides an emulator of system dynamics, enabling offline application for optimal control. This approach not only removes model bias, but also  offers an understanding of the underlying physics, making it superior to strictly whitebox or blackbox approaches.

Here, we report on the control of a qubit coupled to Random Telegraph Noise (RTN), a non-Markovian and non-Gaussian noise, with coupling strength ranging from weak to ultra strong. We demonstrate that our method achieves high-fidelity preparation of a universal set of quantum gates as well as a random set of unitaries over all ranges of coupling strength, including strong-coupling regimes where WB models fail.
This method can be applied to any open finite-dimensional quantum system subject to arbitrary classes of noise (including the case of a quantum bath \cite{youssry2022multi, youssry2020characterization} ) and coupling strength.

\section{Methods}
\label{sec:methods}
\subsection{Problem Setting}
\label{subsec:prob  lem}

\begin{figure}
    \centering
    \includegraphics[scale=0.75]{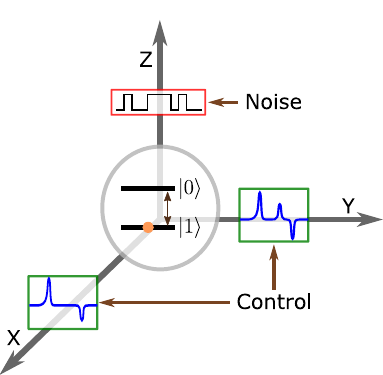}
    \caption{\textbf{The model studied in the paper.} A qubit is subjected to a random telegraph noise along the Z-axis. Control pulses are applied on the X- and Y- axes, and the target is to implement a desired quantum gate.}
    \label{fig:1}
\end{figure}

We investigated a one-qubit system, as shown in \Cref{fig:1}, driven by time-dependent control in the presence of \emph{random telegraph noise} (RTN), which is a non-Markovian non-Gaussian noise model \cite{dong2023resource}. In this context, the non-Markovianity refers to the fact that the noise exhibits non-trivial temporal correlations (colored noise) which induce the non-Markovianity of the system's dynamics. The system is described by the following Hamiltonian:
\begin{equation}
    H(t) = H_{\text{ctrl}}(t) + H_1(t),
\end{equation}
where the control is introduced via
\begin{equation}
    H_{\text{ctrl}} = f_x(t) \sigma_x + f_y(t) \sigma_y
    \label{equ:Hctrl}
\end{equation}
and the coupling of the system to the noise is given by \begin{equation}
    H_1 = g \cos(\Omega (t-t_0) + \phi) \beta(t) \sigma_z,
\end{equation}
with $g$ the coupling strength to a stochastic process $\beta(t)$ further modulated by a periodic function. The noise $\beta(t)$ is chosen to be an RTN process with switching rate $\gamma$. The coupling is further modulated by a periodic function with unknown frequency $\Omega$ (we will consider both the zero and non-zero cases) and phase $\phi.$ Since the $\Omega$ and $t_0$ are unknown, this amounts to $\phi$ being a random phase uniformly distributed within the range of $[0, 2\pi]$ picked for every run of an experiment (see Supplementary Note 1). This Hamiltonian describes many physical systems, such as  a superconducting qubit driven at resonance, and subjected to dephasing noise due to random qubit frequency fluctuations \cite{superconducting2022}.

The system evolves over the time interval $[0,T]$, and is then measured at time $t=T$. The expectation value of an observable $O$ at time $T$ starting an initial state $\rho$ can then be expressed as \cite{youssry2020characterization}
\begin{equation}
    \langle O(T) \rangle = \Tr[V_O(T)U_{\text{ctrl}}(T) \rho U_{\text{ctrl}}(T)^{\dagger}O].
\end{equation}
Here, $U_{\text{ctrl}}(t) = \mathcal{T} e^{- i \int_0^t  H_{\text{ctrl}}(s) ds},$ and the noise operator $V_O(T)$, which depends on the control, captures all the information about the environment and how it affects the evolution of the system. Thus, $V_O = I_S$ amounts to noiseless evolution.



\subsection{Whitebox models}

Exact analytical expressions for the $V_O$ operator exist only in very special cases~\cite{Paz_Silva_2017}, and generally one requires perturbation techniques to make it tractable. The faithfulness of these perturbative approaches  depends on the coupling of the system to the noise source $g$ and the evolution time $T$. We recognize then two types of whitebox models.

\textit{(I) Closed-system whitebox (CS-WB)}. As a first approach we consider a model that assumes that the evolution of the system is ruled by a system-only Hamiltonian, say $H_S = \sum s_\alpha \sigma_\alpha$, and that the evolution is noiseless, i.e., $g=0.$ In terms of our model, where we used $H_S =0$ for simplicity, this implies that the evolution of the qubit is solely generated by the control Hamiltonian. Generally, this strategy yields good performance if the noise is extremely weak relative to the total evolution time, i.e., $ 0 \sim g T \ll 1,$  as demonstrated in \cite{Yang_2019}.

\textit{(II) Open-system whitebox (OS-WB)}. A more faithful model should include the noise to some extent. To do this we use a perturbative Dyson expansion (See Supplementary Note 2), although others could equally be used. In this language, the time-dependent expectation value of an observables $O$ can be approximated by the $N^{\text{th}}$ order expansion
\begin{equation}
\label{eq:expectation}
    \langle O(T) \rangle = \Tr\left[\sum_{k=0}^{N}\frac{\mathscr{D}_{O}^{(k)}(T)}{k!}\rho_s \Tilde{O}(T) \right],
\end{equation}
where $\tilde{O}(T) = U_{\text{ctrl}}^{\dagger}(T) O  U_{\text{ctrl}}(T)$, and the $k$-th Dyson term
\begin{equation}
    \frac{\mathscr{D}_{O}^{(k)}(T)}{k!} = (-i)^k \int_{-T}^{T} d_{>}\vec{t}_{[k]}\langle \Tilde{H}_O(t_1) \ldots \Tilde{H}_O(t_k)\rangle,
\end{equation}
is defined in terms of the effective Hamiltonian,
\begin{align*}\tilde{H}_O(t) =  \begin{cases}
        -\Tilde{O}^{-1}(T)H_I(T-t)\Tilde{O}(T), \quad & 0 \le t \le T\\
        H_I(T+t), \quad &-T \le t < 0,
    \end{cases}
\end{align*}
and we introduced the notation $\int_{-T}^{T}d_{>}\vec{t}_{[k]} = \int_{-T}^{T}dt_1\int_{-T}^{t_1}dt_2 \ldots \int_{-T}^{t_{k-1}}dt_k$. Importantly, the $k$-th order term necessarily depends on the correlation function $\langle \beta(t) \cdots \beta(t_k) \rangle,$ and thus 
$$
\Theta = \{ \langle \beta(t_1) \cdots \beta(t_k) \rangle\}_{k=1}^N.    
$$
For the RTN model we are considering the leading correlators are given by 
\begin{align*}
\langle \beta(t_1) \beta(t_2) \rangle &= g^2e^{-2\gamma(t_1-t_2)},\\
\langle \beta(t_1) \cdots  \beta(t_4) \rangle& =g^4e^{-2\gamma(t_1-t_2+t_3-t_4)},
\end{align*}
with odd-order correlators vanishing. Notice also that while it is formally possible to write the expansion to any order, going to larger values of $N$ quickly becomes numerically challenging given the time-ordered integrals involved. Thus, for this whitebox model to be effective, one requires that 
$(g T)^k \sim 0 $ for $k > N$ and that $N$ is not very large.

\subsection{Categorisation of Coupling strengths}

\begin{figure}
    \centering
    \includegraphics[]{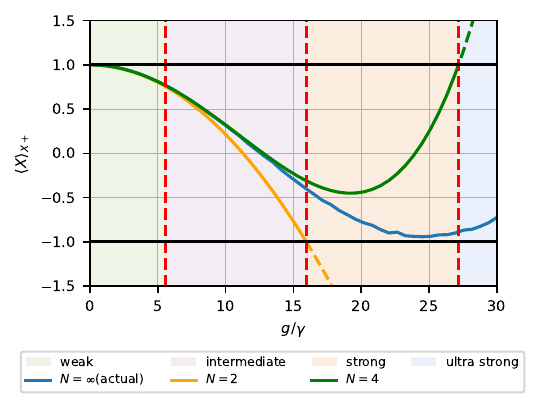}
    \caption{\textbf {The qubit coherence under RTN vs the coupling strength between system and noise.} The blue line shows the actual calculation using Monte-Carlo simulation. The green and orange lines show the theoretical calculation using the Dyson expansion with truncation orders $N = 2,4$ respectively. The dotted red lines define the boundaries between the different noise strength regions.}
    \label{fig:2}
\end{figure}

To get an idea of when this expansion is a faithful representation of the dynamics in our model, we compare the expectation values of a relevant observable using a Monte-Carlo simulation and a truncated Dyson expansion. Concretely, \Cref{fig:2} shows the coherence as a function of $g/\gamma,$ obtained by truncating \Cref{eq:expectation} to order $N = 2, 4$ (See Supplementary Note 2) and using the full simulation (denoted by $N =\infty$) for the case of no control. The coherence is defined as the expectation of the Pauli $X$ observable with initial state $\ket{+}$ over a fixed time $T$. This allows us to heuristically classify the coupling strength into four regions: weak, intermediate, strong, and ultra strong. In the {\it weak} coupling region, the expectation values from both truncation orders closely align with each other (and to the actual values). When the distance is larger than some fixed threshold, say $\epsilon = 0.01$, we transition to the {\it intermediate} coupling regime. This regime extends until where the second order expansion becomes unphysical, i.e., expectation values exceed the range of $[-1,1]$, at which point we move to the {\it strong} regime. Similar criteria is applied to the transition to the {\it ultra strong} regime, but for the $N=4$ expansion. It should be noted that control can drastically alter these definitions due to its noise suppressing effect, but we chose free evolution as the worst-case benchmark. Similar plots for the modulated RTN which can be found in Supplementary Figure 1. We use this classification to ensure that we study the problem across a wide range of coupling regimes, but we highlight that the success of the GB by no means depends on this choice. 

To conclude this section, we stress that the GB and WB model proposed in this paper can apply generally, and the RTN model chosen allows us to directly compare them in a relevant setup. Indeed, other noise models have been previously explored \cite{youssry2021noise, youssry2022multi}. To make this comparison more dramatic, we shall assume that the whitebox models are capable of acquiring perfect information, and thus outperforms a data-driven whitebox as in QNS methods. 

\subsection{Graybox model}

The GB model generally combines whiteboxes and blackboxes. The inputs to our GB model are the control pulses $\Theta$ and the outputs are the expectation of the system's observables given arbitrary initial states. We use the Pauli operators $O \in \{X, Y, Z\}$ as obswrvables, and Pauli eigenstates $\rho \in 
\{\alpha_s\}_{\alpha,s}$, with $\alpha_s$ the $s$ as the initial states. 

In this case, we can obtain a GB model by simply replacing the perturbative equations in the OS-WB model by a blackbox. Concretely, for our blackbox we use two layers of Gated Recurrent Units (GRU) with 60 hidden nodes. This part is utilized to learn the parameters of the $V_O$ operators for each of the three observables as a function of the control pulses. We construct datasets to train and test our model. Each instance in the training dataset is comprised of a pair consisting of a control pulse sequence and the corresponding set of output expectation values. In this paper, we use simulated datasets, but for an experimental application, those datasets will be constructed from measured expectation values. 
More details on the construction of the model and the training protocol can be found in \cite{youssry2020characterization,youssry2021noise,youssry2022multi}.

\subsection{Optimal control}

The key element for any optimization is the cost function $C(u, \Theta; G),$ here defined in terms of the expectation values of observables at the time $T$. Concretely, we will use the MSE between the ideal effect of a gate $G$ on the observables and the actual effect given a control parameterized by $u$. In other words, we will optimize the function
\begin{align}
\label{eq:mse}
    C(u, \Theta; G) = \sum_{\rho,O} \left(\Tr{\left[G \rho G^{\dagger} O\right]} - \Tr{\left[\hat{\rho}(u, \Theta, T) O\right]}   \right)^2, 
\end{align}
where $\rho$ is the initial state, $O$ is an observable, and $\hat{\rho}(u, \Theta, T)$ is the simulated final state at time $T$ using control parameters $u$ in the corresponding model, i.e., CS-WB, OS-WB or GB with model parameters $\Theta$. We choose $\rho$ and $O$ to be informationally-complete sets, where for a single qubit we use the Pauli operators and eigenstates $O \in \{X, Y, Z\}$ and $\rho \in 
\{\alpha_s\}_{\alpha,s}$, 
with $\alpha_s$ the $s$ eigenstate of the Pauli $\sigma_{\alpha}$. We utilize ADAM optimizer \cite{Adam} to find the optimal control pulses. Our strategy involves parameterizing the control signal and computing the gradient with respect to these parameters, as opposed to optimizing the gradient at each time step as done in GRAPE \cite{khaneja2005optimal}. The gradient is calculated through the use of Tensorflow autodifferentiation \cite{keras, tensorflow}. 

It is worth noting that the optimum will depend on the actual cost function used. The cost function defined above is amenable to our algorithms, but other functions could also be used. The process matrix fidelity \cite{wood2015tensor} for our optimal solutions, is defined by the Hilbert-Schmidt distance between the target process matrix $\chi_{\text{target}}$ and actual process matrix $\chi_{\text{actual}}$ obtained from the simulation (see Supplementary Note 3),
\begin{equation}
\label{eq:fidelity}
    \text{Fid}(\chi_{\text{actual}},\chi_{\text{target}}) = \Tr(\chi_{\text{actual}}^{\dagger} \chi_{\text{target}}).
\end{equation}

\begin{figure*}
    \centering
    \includegraphics[width=0.9\textwidth]{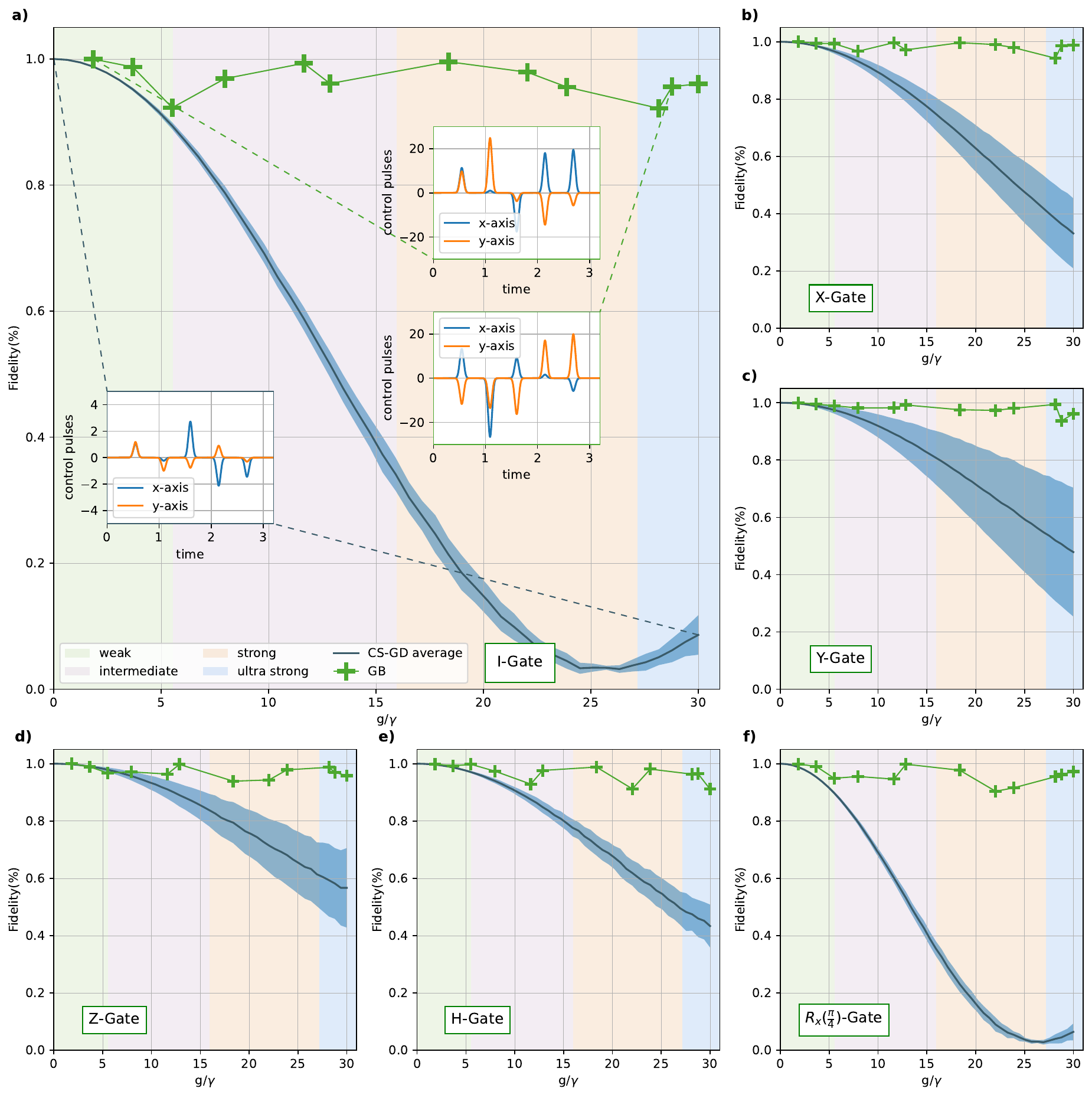}
    \caption{\textbf{The gate fidelity results for RTN case.} Fidelity comparison across $g/\gamma$ values for different gates: a) I, b) X, c) Y, d) Z, e) H, f) $R_x(\frac{\pi}{4})$. For each level of noise strength, three points in the $g/\gamma \in [0,30]$ are selected for the GB approach. In the case of the CS-WB approach, 10 optimal pulses are applied on the noiseless system to simulate performance across all $g$ values. The results are shown in the blue shade. The black line inside of the blue shade represents the average fidelity. In a) the pulse sequence obtained using the CS-WB method is reported. This sequence is used for all values of noise strength. For the GB control, two examples are reported for $g/\gamma$ = 1.8 and 28.8.}
    \label{fig:3}
\end{figure*}

\section{Results}
\label{sec:experiments}

{\it Noisy dynamics simulations.}  We developed a noisy qubit simulator that simulates the dynamics of the qubit using the Monte-Carlo method with parameters: $T = 3.2 \text{ } \mu \text{S} $, number of discrete time steps = 3000, and number of noise realizations = 2000. The implementation is carried out using Python and TensorFlow \cite{keras, tensorflow}. 
Our experiments are conducted under three different noise models: RTN (unmodulated), modulated RTN with $\Omega = 0.5 \text{ MHz}$, and modulated RTN with $\Omega = 1 \text{ MHz}$, all with $\gamma = 0.02 \text{ MHz}$.

{\it Control constraints.} We chose the control constraints such that $f_x(t)$ and $f_y(t)$ in \Cref{equ:Hctrl} are  Gaussian-shaped pulses with fixed positions but variable strength. The waveform $f_{\alpha}(t), \alpha \in \{x,y\}$ is defined as
\begin{equation}
    f_{\alpha}(t) = \sum_{k=1}^{N_P} A_{k,\alpha} e^{-(t-\tau_k)^2/\sigma^2},
\end{equation}
where $N_p=5$ is the number of pulses in the sequence, $A_{k,\alpha}$ is the amplitude of the $k^{\text{th}}$ pulse along the $\alpha$-direction, $\tau_k = \frac{k}{N_p+1}T$ is the location of the pulses, and $\sigma$ is the pulse width chosen to minimize the overlap between the pulses \cite{Perrier_2022}.

{\it Graybox training.} 
With the GB approach, we assembled a dataset consisting of 100,000 samples for each coupling strength $g$ in the range $[0,30/\gamma]$. We sampled three points in each noise regime. During the training phase, the model gains proficiency in estimating the $V_O$ given a set of inputs $\Theta$, thereby establishing the system's dynamics based on the dataset. Subsequently, we validated the model's performance. 

{\it Optimization.}
Once the noisy system was identified, we used the model to derive the  optimal control pulses. To assess the performance of both GB and CS-WB, we rerun the Monte-Carlo simulation on the obtained pulses and calculated the process matrix fidelity as per \Cref{eq:fidelity}. \Cref{fig:3} shows the fidelity results across various values of $g/\gamma$ for a universal set which includes the identity $I$, the Pauli $\sigma_x, \sigma_y, \sigma_z$, the Hadamard $H$, and the rotation  $R_X(\pi/4)$ gate in the case of RTN model. For the modulated RTN case with $\Omega = 0.5$ and $\Omega = 1$, see Supplementary Figures 2 and 3. The optimal pulses derived from both the CS-WB and GB approaches for the RTN model at $g/\gamma=28.8$ can be found in Supplementary Material Figures 4 and 5.  In the CS-WB approach, we performed 1000 iterations for optimization. Subsequently, we generated ten optimal solutions for each gate, employing these optimized pulses across 50 distinct coupling strengths $g$ within the range of $[0, 30/\gamma]$ to span from weak to ultra strong coupling. The blue-shaded areas in the plots of \Cref{fig:3} show the distribution of fidelity over the different solution obtained from the CS-WB model.

\begin{figure*}
    \centering
    \includegraphics[width=\textwidth]{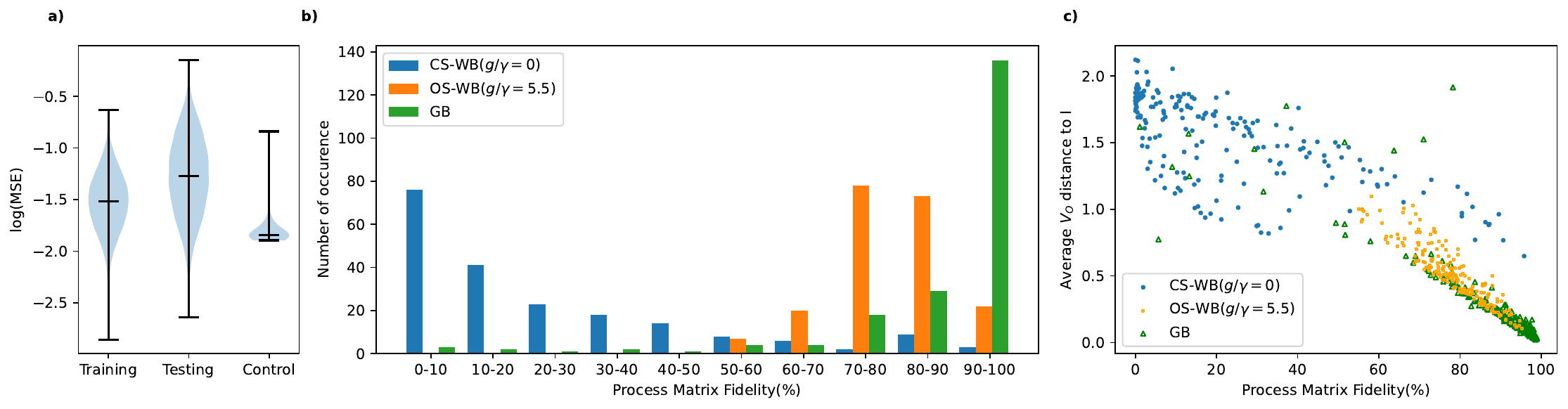}
    \caption{\textbf{Quantum control results for \boldmath$\boldsymbol{g/\gamma = 30}$. } a) A violin plot displaying MSE statistics of GB control for both training and testing datasets. Each example in the dataset is compared against the predictions of the GB, and the MSE is then computed. The lower, middle and upper horizontal lines indicate the minimum, median, and maximum values respectively. b) GB control of $200$ random unitaries for a system  under RTN with $g/\gamma = 30$, compared to CS-WB approach on a noiseless system and  CS-WB approach on a system under RTN with $g/\gamma = 5.5$, which is the limit for the truncated Dyson expansion to be physical. The optimized pulses from all three methods are then simulated for a system with $g/\gamma = 30$, and the process matrix fidelity is calculated. The GB method outperforms gradient descent, with over $85\%$ of unitaries achieving fidelity greater than $80\%$. c) The average distance between $V_O$ operator and $I$ for $200$ random unitaries. GB shows distance less than $0.5$ for unitaries with fidelity higher than $70\%$. }
    \label{fig:4}
\end{figure*}

In \Cref{fig:4}a we report violin plots of the MSE for the GB training, testing and control for the system under RTN with $g/\gamma = 30$, where the GB achieves a mean MSE of  0.036 and 0.077 for the training and validation data respectively. \Cref{fig:4}b, shows the fidelity distribution of $200$ randomly-generated unitaries for a the case of $g/\gamma=30$ noise. We used the trained GB model to find the optimal pulses and compared them against those obtained from the CS-WB model and the OS-WB model with $g/\gamma = 5.5$.  
For GB, around $70\%$ of unitaries achieve more than $90\%$ and over $85\%$ achieving fidelity above $80\%$. This outperforms all other approaches. 

To further understand the performance of the different models, we calculated the average Frobenius distance between the noise $V_O$ and the identity $I$ operators, $\frac{1}{3}\sum_{O \in X,Y,Z} ||V_O-I||_F$, where the operators are evaluated at the optimal solution. The noise operator can be calculated using a Monte-Carlo simulation \cite{Perrier_2022,youssry2020characterization}. \Cref{fig:4}c shows that the average distance is strongly correlated to the process fidelity, i.e. the control solutions with high fidelity have $V_O$ operators closer to the identity. This is expected since the $V_O$ captures the deviation from closed-system dynamics, i.e. unitary evolution. The GB achieves smaller distances than whitebox approaches, effectively cancelling the noise.

\section{Discussion}
\label{sec:discussion}
In this paper, we have reported on a graybox quantum control approach for quantum systems subjected to an unknown strongly-coupled noise, where the standard control strategies fail. We applied our method to non-Markovian, non-Gaussian noise, a critical a challenge for quantum technology, particularly in the strongly-coupled regime, where currently-known whitebox methods become intractable. Single-axis RTN noise model was chosen to enable the computations of the WB methods for comparison purposes. However, our GB method is not restricted to a particular noise type or strength and can be applied multi-axis and quantum-bath noise \cite{youssry2022multi}.

Our work highlights the limitations of the whitebox approach, where the faithfulness of the fixed perturbation order models is not guaranteed in the strong-coupling regime. In the GB model, the ``black'' part addresses the challenge of expressing the dynamics in closed form, which is required for a fully whitebox model, rendering the method suitable for strong-coupling regime. Our results complement the recent work on the superiority of GB over standard blackbox approaches \cite{perrier2020quantum, youssry2023experimental}. 
Furthermore, a trained GB model can be used for any desired gate, whereas model-free methods such as reinforcement learning require re-training for each target operation. This positions the GB as the preferred approach for the control of noisy quantum systems. 

As depicted in \Cref{fig:3}, the GB approach, without any assumptions about the underlying noise, achieved higher than $90\%$ fidelity across the entire spectrum of coupling strengths. It consistently outperformed the CS-WB model, with a widening performance gap as the noise strength increases. The control fidelities obtained using the CS-WB approach had also a wide distribution. Even though they are all optimal in the weak-coupling $g/\gamma \approx 0$, the variations steadily increase as the noise increases, up to $50\%$ at $g/\gamma=30$ in the worst case for the $Y$-gate. In \Cref{fig:4}, we explored the ultra strong coupling regime ($g/\gamma = 30$). Testing the GB on 200 random unitaries against the CS-WB approach and OS-WB approach with $g/\gamma = 5.5$, over $85\%$ of the random unitaries achieved fidelity exceeding $80\%$, surpassing the performance of both whitebox methods. 

In our experiments, we observed that MSE of the training underestimates the performance of the GB in terms of control fidelity. This observation aligns with findings from Dong et al.'s study \cite{dong2023resource}. \Cref{fig:4}a illustrates that while the GB MSE statistics for both training and testing datasets under RTN with $g/\gamma = 30$ are not notably low, the control achieved over $90\%$ fidelity. By comparing results across different scenarios—untrained model, $10,000$ dataset examples, and $20,000$ dataset examples—we found that the fidelity increases with the number of examples in the dataset.

To enhance the graybox approach, the dataset size and/or  the amplitude of the control pulses can be increased. In general, the stronger the noise, the larger the required dataset. Our experiments improved performance when the dataset samples were increased from 10,000 to 100,000. However, the architecture of the GB remains the same, as opposed to the whitebox methods, where more complicated expressions are needed. Addressing the dependence on large datasets is crucial, and requires further investigation. Another aspect that can be considered in the future, is to investigate the performance of deep quantum circuits subjected to non-Markovian noise, which is effectively a strongly-coupled system, in comparison to Markovian noise.

\noindent\textbf{Acknowledgements} AP acknowledges an RMIT University Vice-Chancellor’s Senior Research Fellowship and a Google Faculty Research Award. This work was supported by the Australian Government through the Australian Research Council under the Centre of Excellence scheme (No: CE170100012).  Work at Griffith University was supported (partially) by the Australian Government via AUSMURI Grant (No: AUSMURI000002). This research was also undertaken with the assistance of resources from the National Computational Infrastructure (NCI Australia), an NCRIS enabled capability supported by the Australian Government.

\bibliography{references}

\end{document}


\title{Supplementary Materials for 
 \\ Quantum control in the presence of strongly coupled non-Markovian noise}
\author{Arinta Auza}
\address{Quantum Photonics Laboratory and Centre for Quantum Computation and Communication Technology, RMIT University, Melbourne, VIC 3000, Australia}

\author{Akram Youssry}
\address{Quantum Photonics Laboratory and Centre for Quantum Computation and Communication Technology, RMIT University, Melbourne, VIC 3000, Australia}

\author{Gerardo Paz-Silva}
\address{Centre for Quantum Dynamics and  Centre for Quantum Computation and Communication Technology, Griffith University, Brisbane, Queensland 4111, Australia}

\author{Alberto Peruzzo}
\address{Quantum Photonics Laboratory and Centre for Quantum Computation and Communication Technology, RMIT University, Melbourne, VIC 3000, Australia}
\address{Qubit Pharmaceuticals, Advanced Research Department, Paris, France}

\maketitle

\section*{Supplementary Note 1: Noise Models}

We employ random telegraph noise (RTN) in our system model. This is a binary process which switches between states at any given time with probability $\gamma$, so that the switches occurring in $[0, T]$ follow a Poisson distribution with a mean of $\gamma T$. It can be defined by the random process:
\begin{equation}
\xi(t) = \xi(0)(-1)^{n(0,t)}
\end{equation}
Here, $n(0,t)$ represents the number of switches during the time interval $[0,t]$, following a Poisson distribution with $\langle n(0,t)\rangle = \gamma t$. The initial value $\xi(0)=\pm 1$ is chosen randomly by flipping an unbiased coin. The second and fourth-order moments are given by:

\begin{align}
    \langle \xi(t_1)\xi(t_2))\rangle &= e^{-2\gamma(t_1-t_2)} \\
    \langle \xi(t_1)\xi(t_2)\xi(t_3)\xi(t_4) \rangle &= e^{-2\gamma (t_1 - t_2 + t_3 - t_4)}
\end{align}

For a more general case where the RTN noise is with a deterministic modulation frequency $\Omega$, and random phase shift $\phi$ drawn from the uniform distribution $[0,2\pi]$. we define
\begin{equation}
    \beta(t)= \cos (\Omega t + \phi) \xi(t)
\end{equation}
Then, the second and fourth-order moment are given by:
\begin{align}
    \langle \beta(t_1)\beta(t_2) \rangle &=  \cos(\Omega(t_1-t_2))e^{-2\gamma(t_1-t_2)}\\
    \langle \beta(t_1)\beta(t_2)\beta(t_3)\beta(t_4) \rangle &=  [ \cos(\Omega(t_1+t_2-t_3-t_4)) + \cos(\Omega(t_1-t_2+t_3-t_4) \nonumber\\
    &\qquad+\cos(\Omega(t_1-t_2-t_3+t_4))] e^{-2 \gamma (t_1-t_2+t_3-t_4)}
\end{align}

\section*{Supplementary Note 2: Perturbative Dyson Expansion of Observables}

Given the total Hamiltonian of the system,
\begin{equation}
    H(t) = g\beta(t) \sigma_z + f_x(t)\sigma_x + f_y(t)\sigma_y,
\end{equation}
we can move to the interaction picture with respect to the control Hamiltonian $ H_c(t) = f_x(t)\sigma_x + f_y(t)\sigma_y$,
\begin{align}
    U(t) &= \mathcal{T}_{+}\left(e^{-i\int_{0}^{t}H(s)ds}\right) := U_c(t) U_I(t),
\end{align}
where the control unitary is given by
\begin{align}
    U_c(t) &= \mathcal{T}_{+} \left(e^{-i\int_{0}^{t}H_c(s)ds}\right),
\end{align}
and the interaction Hamiltonian in the interaction picture is given by,
\begin{align}    
    H_I(t) &= (U_c^{\dagger}(t) \sigma_z U_c(t)) g \beta(t)\\
    &= g\sum_{a} y_{a}(t)\sigma_a\beta(t)\\
    y_a(t) &= \Tr[U_c^{\dagger}(t)\sigma_z U_c(t) \sigma_a]/2.
\end{align}
Here we use the fact that Pauli operators form an orthonormal set of basis for any $2\times 2$ operator.
We can now express the expectation of an observable $O$ at time $t=T$ starting from the initial state $\rho_s$ as
\begin{align}
    \bracket{O(T)} &= \bracket{\Tr[U(T)\rho_sU^{\dagger}(T)O]}\\
  &=\bracket{\Tr[U_c(T)U_I(T)\rho_sU_I^{\dagger}(T)U_c^{\dagger}(T)O]}\\
    &= \Tr[\bracket{U_I(T)\rho_sU_I^{\dagger}(T)}U_c^{\dagger}(T) O U_c(T)]\\
    &\approx \Tr[\bracket{(\mathbb{I} + D_1 + D_2)\rho_s(\mathbb{I}+D_1^{\dagger}+D_2^{\dagger})}  \Tilde{O}(T)]\\
    &\approx \Tr[\bracket{\rho_s + D_1\rho_s +\rho_sD_1^{\dagger} + D_1\rho_sD_1^{\dagger} + D_2\rho_s + \rho_sD_2^{\dagger}}\Tilde{O}(T)],
\end{align}
where $\tilde{O}(T):=U_c^{\dagger}(T) O U_c(T)]$, and we restrict the Dyson expansion in the last two lines to term up to second order in $g$. The Dyson terms are defined as
\begin{align}
    D_1 &= -i \int_{0}^{T}ds \text{ } H_I(s)\\
    D_1^{\dagger} &= i \int_{0}^{T}ds \text{ } H_I(s)\\
    D_2 &= -\int_{0}^{T}ds\int_0^s ds' H_I(s)H_I(s')\\
    D_2^{\dagger}&= -\int_0^Tds \int_0^s ds' H_I(s')H_I(s)
\end{align}
We can then calculate the terms inside the trace as follows.
\begin{align}
     \Tr[\bracket{D_1\rho_s}\Tilde{O}(T)] &= \left\langle -i\int_0^T ds\sum_a y_a(s)\sigma_a g \beta(s)\rho_s \Tilde{O}(T) \right\rangle\\
    &= -ig \int_0^T ds \sum_a y_a(s) \bracket{\beta(s)} \Tr[\sigma_a \rho_s \Tilde{O}(T)]\\
    &=0,
\end{align}
because we assume the noise has zero mean. Similarly, we have $\Tr[\bracket{\rho_s D_1^{\dagger}}\Tilde{O}(T)] =0 $. Next, we have
\begin{align}
    \Tr[\bracket{D_1\rho_sD_1^{\dagger}}] &= \sum_{a,a'}\int_0^T dr \int_0^{T}dr' y_a(r)y_{a'}(r') \bracket{g\beta(r)g\beta(r')} \Tr[\sigma_a\rho_s\sigma_{a'}\Tilde{O}(T)]\\
    &= g^2\sum_{a,a'} I_{a,a'}\Tr[\sigma_a\rho_s\sigma_{a'}\Tilde{O}(T)],
\end{align}
where the $I_{a,a'}$ can be calculated numerically given the prior knowledge of the second-order correlation function of the noise $ \bracket{\beta(t_1)\beta(t_2)}$. The two remaining terms are
\begin{align}
    \Tr[\bracket{D_2\rho_s}\Tilde{O}(T)] &= -\sum_{a,a'}\int_0^T dr \int_0^{r}dr' y_a(r)y_{a'}(r') \bracket{g\beta(r)g\beta(r')} \Tr[\sigma_a\sigma_{a'}\rho_s\Tilde{O}(T)]\\
    &:= -g^2\sum_{a,a'} I_{a,a'}^{>} \Tr[\sigma_a\sigma_{a'}\rho_s\Tilde{O}(T)]\\
    \Tr[\bracket{\rho_s D_2}\Tilde{O}(T)] &= -\sum_{a,a'}\int_0^T dr \int_0^{r}dr' y_a(r')y_{a'}(r) \bracket{g\beta(r')g\beta(r)} \Tr[\rho_s\sigma_a\sigma_{a'}\Tilde{O}(T)]\\
    &:= -g^2\sum_{a,a'} I_{a,a'}^{<}\Tr[\rho_s\sigma_a\sigma_{a'}\Tilde{O}(T)],
\end{align}
where $I_{a,a'}^{>}$ and $I_{a,a'}^{<}$ can also be calculated numerically.

As a special case, in the absence of control pulses, (i.e. only drift and noise affecting the qubit), the equations can be further simplified. In this case $H_c(t) = 0$, $U_c(t) = \mathbb{I}$, and $\tilde{O}(T) = O$. The interaction Hamiltonian reduces to
\begin{align*}
    H_I(t) = g \beta(t) \sigma_z &\implies y_a(t) = \begin{cases}
        1, \quad a=z\\
        0, \quad \text{otherwise}
    \end{cases}
\end{align*}
The integral terms $I_{a,a'}, I_{a,a'}^>, I_{a,a'}^<$ thus vanish for all indices except for $(z,z)$ which can be calculated as,
\begin{align*}
    I_{z,z} &= \int_0^T dr \int_0^T  dr' \bracket{\beta(r)\beta(r')}\\
    I_{z,z}^> &= \int_0^T dr \int_0^r  dr' \bracket{\beta(r)\beta(r')}\\
    I_{z,z}^< &= \int_0^T dr \int_0^r  dr' \bracket{\beta(r')\beta(r)}.
\end{align*}
For classical noise $\beta(t)$, the correlation function is symmetric with respect to its two arguments. Thus, $I_{z,z'} = 2 I_{z,z'}^> = 2 I_{z,z'}^<$. he Dyson expansion up to second order of $g$ reduces to
\begin{align*}
    \bracket{O(T)} = \Tr[\rho_s O] + g^2 I_{z,z} \left( \Tr[\sigma_z \rho_s \sigma_z O] - \Tr[\rho_s O]\right).
\end{align*}
If we are interested in calculating the coherence of the qubit, then we have $\rho_s = X_+$, $O=X$. Therefore,
\begin{align}
    \bracket{X(T)} &= 1 + g^2 C_2,\\
     C_2 &= -4\int_0^T \int_0^{t_1} \bracket{\beta(t_1)\beta(t_2)}dt_2 dt_1   
\end{align}
The calculations shown here can be extended to higher-order expressions as well. For a noise with a vanishing third-order correlation function (such as RTN noise), the fourth-order expansion is
\begin{align}
    \bracket{X(T)} &= 1 + g^2 C_2 + g^4 c_4,\\
     C_2 &= -4\int_0^T \int_0^{t_1} \bracket{\beta(t_1)\beta(t_2)}dt_2 dt_1   \\
     C_4 &= 16\int_{0}^{T}\int_{0}^{t_1}\int_{0}^{t_2}\int_{0}^{t_3} \bracket{\beta(t_1)\beta(t_2)\beta(t_3)\beta(t_4)}dt_4 dt_3 dt_2 dt_1.
\end{align}
We selected the weak-coupling region based on the criterion that the difference between the second and fourth-order Dyson expansions is smaller than a specified threshold, denoted as $\epsilon$. The transition to the intermediate/strong coupling region occurs when this difference exceeds $\epsilon$, and concurrently, the expectation falls below $-1$. Beyond an expectation value of $+1$, we identify the subsequent region as the ultra-strong region.

Finally, the most general form of the expansion up to arbitrary order $N$ can be calculated as follows. Define
\begin{align*}
       \tilde{H}_O(T) &= \begin{cases}
        -\Tilde{O}^{-1}(T)H_I(T-t)\Tilde{O}(T), \quad & 0 \le t \le T\\
        H_I(T+t), \quad &-T \le t < 0
    \end{cases}
\end{align*}
Then we have,
\begin{equation}
\label{eq:expectation}
    \langle O(T) \rangle = \Tr\left[\sum_{k=0}^{N}\frac{\mathscr{D}_{O}^{(k)}(T)}{k!}\rho_s \Tilde{O}(T) \right],
\end{equation}
and the $k$-th term of Dyson contribution is expressed by 
\begin{equation}
    \frac{\mathscr{D}_{O}^{(k)}(T)}{k!} = (-i)^k \int_{-T}^{T} d_{>}\vec{t}_{[k]}\langle \Tilde{H}_O(t_1) \ldots \Tilde{H}_O(t_k)\rangle,
\end{equation}
where $\int_{-T}^{T}d_{>}\vec{t}_{[k]} = \int_{-T}^{T}dt_1\int_{-T}^{t_1}dt_2 \ldots \int_{-T}^{t_{k-1}}dt_k$. For more details, check \cite{dong2023resource, Paz_Silva_2017}. It is clear that the calculations becomes quickly intractable for higher-order terms. 

\section*{Supplementary Note 3: Process matrix}
We can derive the expectation from \cite{wood2015tensor} $\langle O\rangle = \Tr[\mathcal{E}(\rho)O]$. Let $\chi$ be a process matrix. Then, we can obtain $\mathcal{E}$ from
\begin{equation}
    \mathcal{E}(\rho) = \sum_{\alpha, \beta} \chi_{\alpha, \beta}\sigma_{\alpha}\rho\sigma_{\beta}^{\dagger} 
\end{equation}

Let $\Gamma = \{\sigma_{0}, \sigma_{1}, \sigma_{2}, \sigma_{3}\}$ be a set of Pauli matrices. For initial state $\rho$ and observable $O$, let $E_{ij}$ be the expectation of observable $\sigma_j$ with initial state $\sigma_i$ obtained from the evolution. Then,
\begin{align}
    E_{ij} &= \Tr\left[\sum_{\alpha, \beta} \chi_{\alpha,\beta}\sigma_{\alpha}\sigma_i\sigma_{\beta}^{\dagger}
    \sigma_j\right]\\
    &= \chi_{\alpha,\beta}\sum_{\alpha, \beta} \Tr\left[\sigma_{\alpha}\sigma_{i}\sigma_{\beta}^{\dagger}\sigma{j}\right] \\
\end{align}
Let $\Tilde{\chi}$ be a vectorization of the process matrix $\chi$ such that $\Tilde{\chi}_m = \chi_{m/4,[m]_4}$ and $A$ be a $16$-by-$16$ matrix such that
\begin{equation}
    A_{kl} = \Tr\left[ \sigma_{k/4}\sigma_{l/4}\sigma_{[k]_4}\sigma_{[l]_4} \right]
\end{equation}
Then, if $\Tilde{E}$ is a column vector such that $\Tilde{E}_{m} = E_{m/4,[m]_4}$, we have $\Tilde{E} = A\Tilde{\chi}$. Thus, we can obtain the process matrix $\chi$ from
\begin{equation}
    \Tilde{\chi} = A^{-1}\Tilde{E}
\end{equation}



\section*{Supplementary Figures}
\begin{figure}[H]
    \centering
    \includegraphics{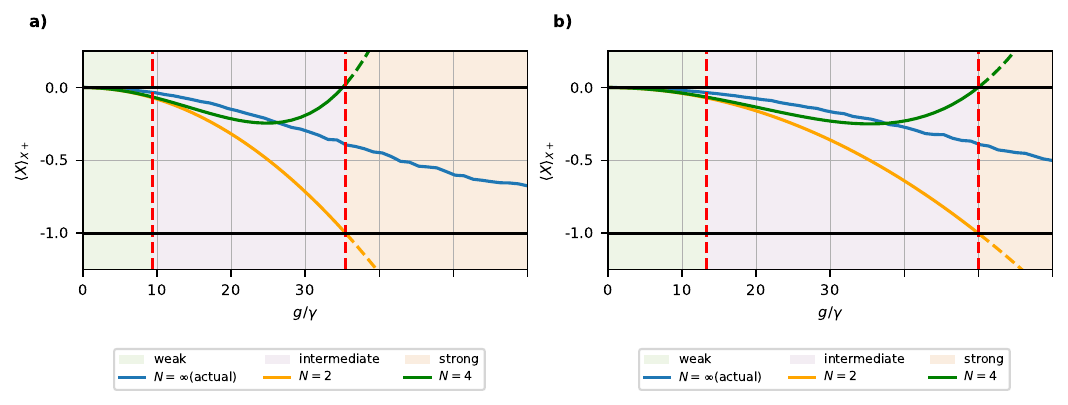}
    \caption{\textbf {The qubit coherence under modulated RTN vs the coupling strength,} with modulation frequency: a) $\Omega = 0.5$, b) $\Omega=1$}
    \label{fig:dyson_mod}
\end{figure}

\begin{figure}[H]
    \centering
    \includegraphics[width=0.9\textwidth]{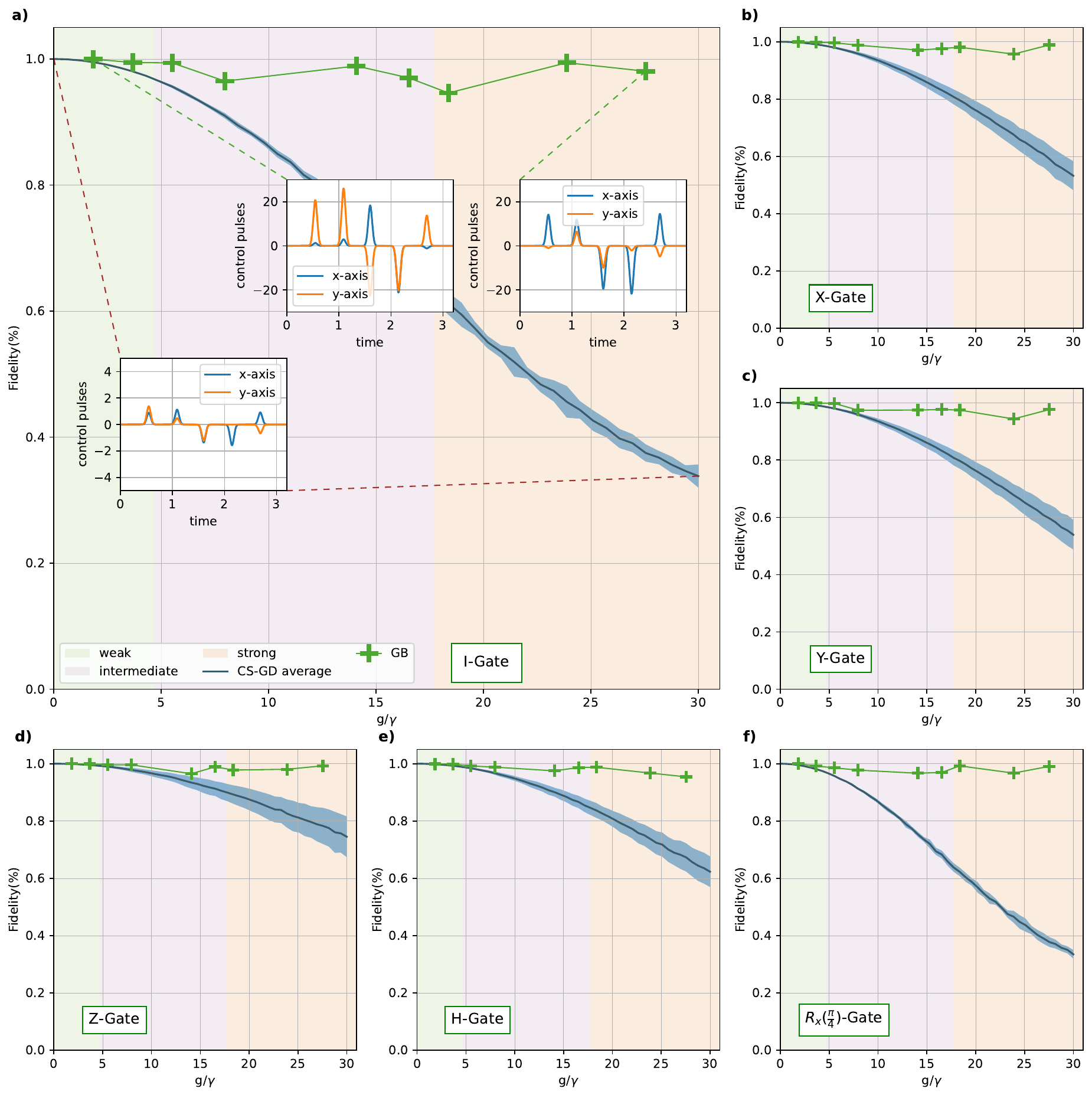}
    \caption{\textbf{The results for modulated RTN with $\Omega=0.5$.} Fidelity comparison across $g/\gamma$ values for different gates: a) $I$, b) $X$, c) $Y$, d) $Z$, e) $H$, f) $R_X(\frac{\pi}{4})$.}
    \label{fig:rtn_mod05}
\end{figure}

\begin{figure}[H]
    \centering
    \includegraphics[width=0.9\textwidth]{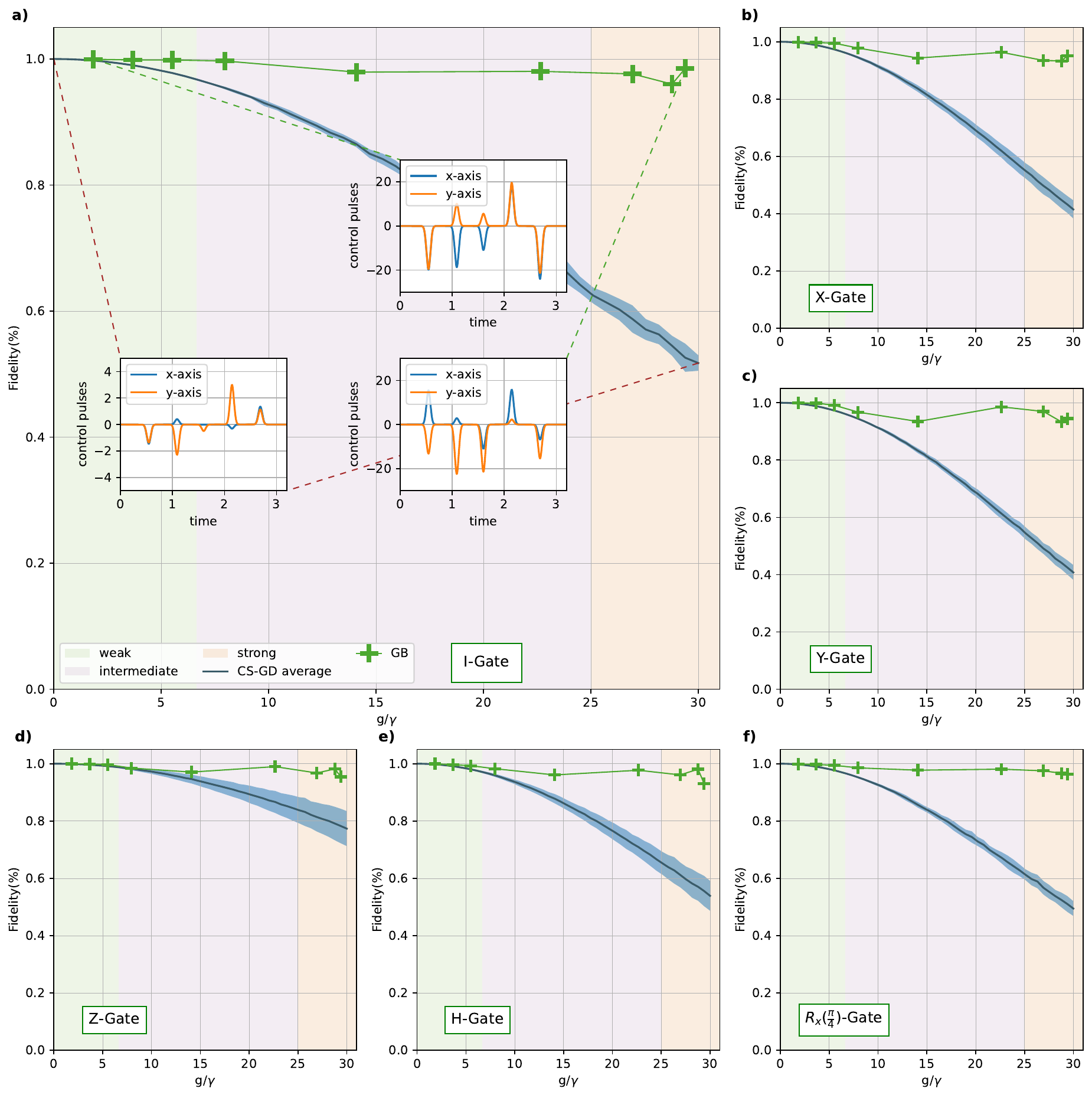}
    \caption{\textbf{The results for modulated RTN with $\Omega=1$.} Fidelity comparison across $g/\gamma$ values for different gates: a) $I$, b) $X$, c) $Y$, d) $Z$, e) $H$, $f$) $R_X(\frac{\pi}{4})$.}
    \label{fig:rtn_mod1}
\end{figure}

\begin{figure}[H]
    \centering
    \includegraphics[width=0.9\textwidth]{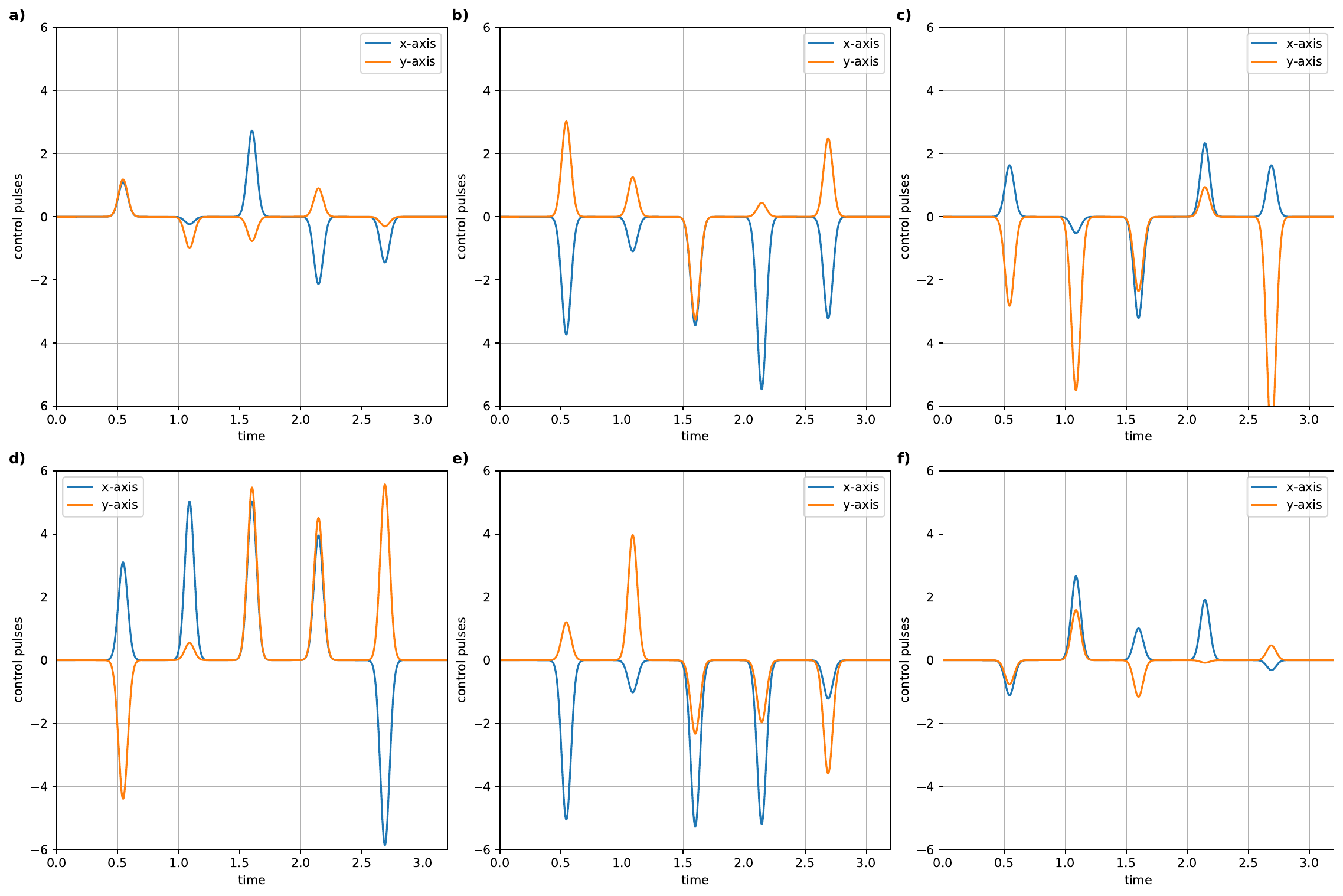}
    \caption{\textbf{Closed-System Whitebox optimized pulses for the RTN model,} for different gates: a) $I$, b) $X$, c) $Y$, d) $Z$ e) $H$, f) $R_X(\frac{\pi}{4})$}
    \label{fig:pulses_gd}
\end{figure}

\begin{figure}[H]
    \centering
    \includegraphics[width=0.9\textwidth]{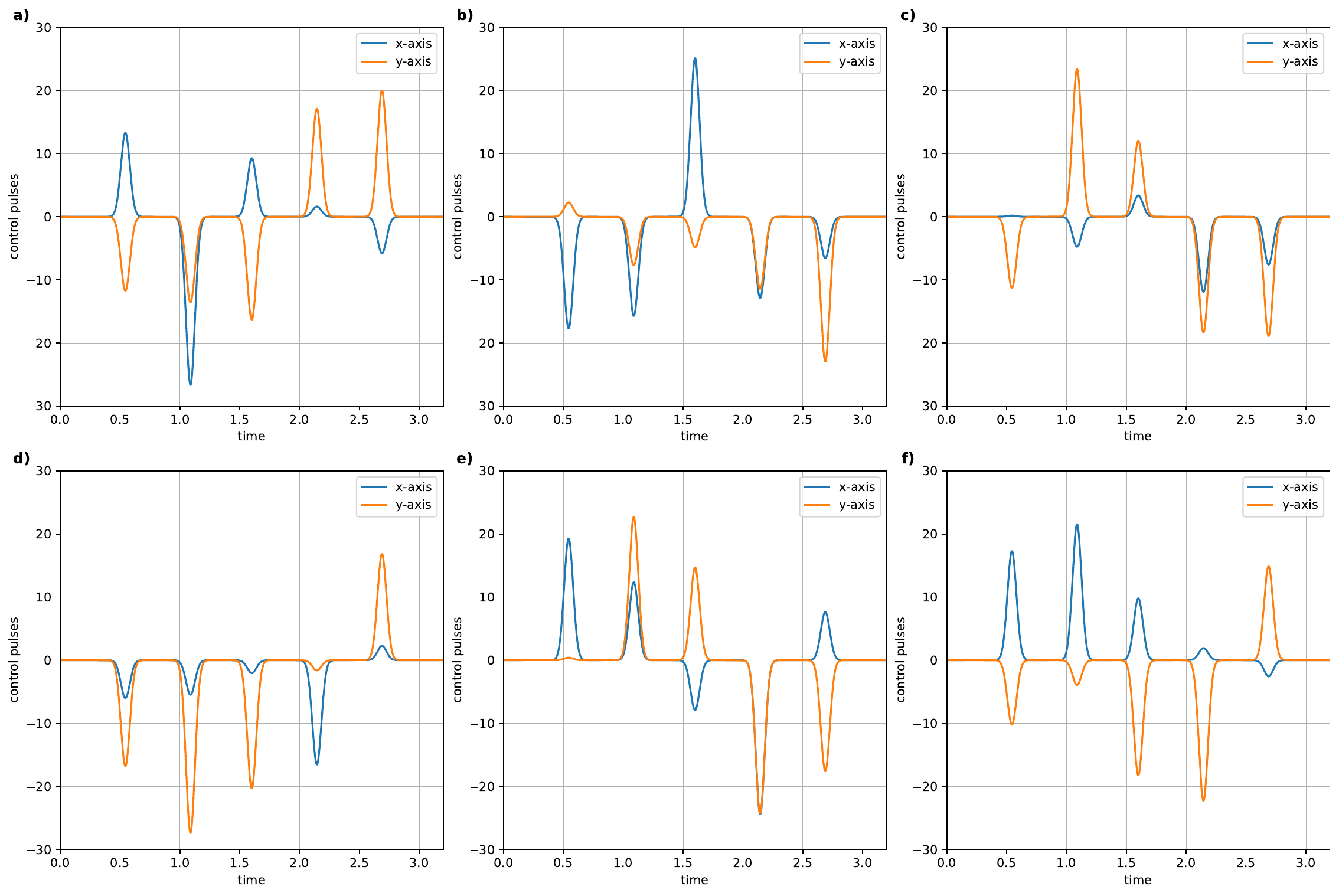}
    \caption{\textbf{Graybox optimized pulses for the RTN model with $g/\gamma = 28.8$,} for different gates: a) $I$, b) $X$, c) $Y$, d) $Z$, e) $H$, f) $R_X(\frac{\pi}{4})$}
    \label{fig:pulses_gb}
\end{figure}

\bibliography{references}